\documentclass[aps,prd,a4paper,twocolumn,showpacs,showkeys,preprintnumbers,amsmath,amssymb]{revtex4}
\usepackage{graphicx}
\usepackage{bm}
\usepackage{dcolumn}

\newcommand{\hide}[1]{{}}
\newcommand{\mr}[1]{\mathrm{#1}}

\newcommand{\cx}{\ensuremath{\mathbb C}}         
\newcommand{\id}{\textrm{1\kern-.25em I}}        

\newcommand{\HH}{\mathcal{H}}

\newcommand{\bx}{\mathbf{x}}
\newcommand{\bk}{\mathbf{k}}

\newcommand{\de}{\delta}

\newcommand{\dd}{\partial}
\newcommand{\ra}{\rightarrow}
\newcommand{\al}{\alpha}
\newcommand{\om}{\omega}
\newcommand{\Om}{\Omega}

\newcommand{\lsim}{\,\raisebox{-0.6ex}{$\buildrel < \over \sim$}\,}

\newcommand{\be}{\begin{equation}}
\newcommand{\ee}{\end{equation}}
\newcommand{\ben}{\begin{equation*}}
\newcommand{\een}{\end{equation*}}
\newcommand{\bea}{\begin{eqnarray}}
\newcommand{\eea}{\end{eqnarray}}
\newcommand{\bean}{\begin{eqnarray*}}
\newcommand{\eean}{\end{eqnarray*}}

\begin{document}

\title{Graviton production in non-inflationary cosmology}

\author{Ruth Durrer and Massimiliano Rinaldi}
\affiliation{Universit\'e de Gen\`eve, D\'epartment de Physique Th\'eorique,
        24 quai Ernest Ansermet, CH--1211 Gen\`eve 4, Switzerland}

\date{\today}

\begin{abstract}
\noindent We discuss the creation of massless particles in a Universe, which 
transits from a radiation-dominated era to any other expansion law. We calculate
in detail the generation of gravitons during the transition to a 
matter-dominated era. We show that the resulting gravitons generated in the
standard radiation/matter transition are negligible. We use our result to 
constrain one or more previous matter-dominated era, or any other expansion 
law, which may have taken place in the early Universe. We also derive a 
general formula for the modification of a generic initial graviton spectrum 
by an early matter dominated era.
\end{abstract}


\maketitle

\section{Introduction}
\noindent One of the most interesting aspect of inflation is that it leads to 
the generation of a scale-invariant spectrum of scalar
perturbations~\cite{slava} and of gravitational waves~\cite{staro} (see
also \cite{slavaBook,myBook}). The origin of these perturbations is the 
quantum generation of field correlations in a time-dependent background
 (scalar field modes for the scalar perturbations and gravitons
for the tensor perturbations).
Since this generation takes place mainly on super-horizon scales, it is not 
correct to talk of 'particles'. However, long after inflation, when the 
perturbations re-enter the horizon, the particle concept, e.g. for gravitons 
becomes meaningful and we can calculate, e.g. the energy density of the 
gravitons which have been generated during inflation.

It is  natural to ask whether particle production takes place also in an 
ordinary expanding but non-inflationary Friedmann Universe.
The answer is that particle production (or more generically the quantum 
generation of field correlations) can indeed take place
after inflation, but if there is no inflationary phase to start with,
the initial vacuum state is in general not known, and the production rate 
cannot be computed. Examples where particle
creation taking place after inflation (or after pre-big bang) modifies
the final spectrum are given in Refs.~\cite{pbb1,pbb2,VS2,VS3}.
Especially in Ref.~\cite{VS2} it has been studied how inflationary 
perturbations are modified if the subsequent expansion is not standard 
radiation but some other expansion law.

In general, the vacuum state, and hence the particle concept, is well 
defined only if the
spacetime is static or very slowly varying~\cite{BirDav}. 
Let us consider a mode of fixed (comoving) frequency $k$ in a Friedmann 
universe. The above  condition then corresponds to $k/\HH\gg 1$, where 
$\HH=aH$ is the comoving Hubble scale.
In this sense, the scale (wavelength) under consideration must 
be ``inside the horizon". However, the production of a particle with a given 
energy $k$ can only take place if the energy scale of expansion is larger 
or of the order of the energy of the associated mode, i.e. $k \lsim \HH$. 
Therefore, having a well defined
initial vacuum state, and subsequent particle creation, usually requires a
decreasing comoving Hubble rate. This is verified only during inflation or
during a collapsing Friedmann Universe, like in the pre-big bang or in bouncing
models.

However, one important exception to this general rule exists,  and it is the 
subject of
the present paper: in a radiation-dominated Friedmann background, massless 
perturbations do not couple to the expansion of the Universe, and evolve like 
in ordinary Minkowski space. This has already been realized and studied to 
some extent in Ref.~\cite{grish}. In a radiation-dominated Universe we therefore
can provide vacuum initial conditions for all modes of a massless field, 
including super-horizon modes. Thus, when the expansion law changes, e.g. 
from radiation- to matter-dominated, the massless modes couple to the 
expansion of the Universe, and those with $k/\HH <1$ are amplified. 

In this paper we study this phenomenon in two situations of interest. In the 
first, we investigate graviton
production during the ordinary radiation/matter transition at redshift
$z_\mr{eq}\simeq 3500$. We determine the amplitude and the spectrum of
the generated gravity wave background, and we show that
the spectrum is flat and the amplitude is negligibly small.
In the second case, we investigate the production
of gravitons during an arbitrary matter-dominated phase, which could take place
in the early Universe, e.g., if a (very weakly interacting) particle becomes
massive, and succeeds to dominate the Universe for a period of time
before it decays into radiation. We derive a general formula for the
gravity wave spectrum generated by any number of such intermediate periods
of matter domination. We also determine the gravity waves produced 
by a transition into an arbitrary other expansion law. Finally, we 
discuss the modifications of our results which occur when the initial state is
not the vacuum but some arbitrary state which may already contain particles.
In this work we concentrate on graviton production, but all our 
results are equally applicable to other massless particles.

The reminder of this paper is organized as follows. In the next section we
present the setup and the basic formulae used in our work. In Section~III, we 
calculate the gravitational wave production during the standard 
radiation/matter transition. We also give the results for the transition 
from radiation to some generic expansion law. In Section~\ref{more} we consider
the effect of one or several additional transitions in the early Universe and
we derive results for general non-vacuum initial conditions. In
Section~\ref{conc} we derive the consequences of our results and we
draw some conclusions.

\noindent{\bf Notation:} We work in a spatially flat Friedmann Universe,
and we denote conformal time by $t$, so that
$$ds^2 = a^2(t)\left(-dt^2 +\de_{ij}dx^idx^j\right)~.$$
An over-dot denotes the derivative with respect to the conformal time.
We use natural units $c=\hbar=1$, except for Newton's constant $G$, which is
related to the reduced Planck mass by $8\pi G =m_p^{-2}$. We normalize
the scale factor, so that $a_0=a(t_0)=1$  at the present time.

\section{Graviton creation in cosmology}\label{zero}

We now consider tensor perturbations of the Friedmann metric, namely
$$ds^2 = a^2(t)\left[-dt^2 +(\de_{ij}+2h_{ij})dx^idx^j\right]~,$$
where $h_{ij}$ is a transverse and traceless tensor. In Fourier space 
we have 
\be
h_{ij}(\bk,t)= h_+(k,t)e^{(+)}_{ij}(\hat\bk) +  h_-(k,t)e^{(-)}_{ij}(\hat\bk)\ ,
\ee
where $e^{(\pm)}_{ij}(\hat\bk)$
denote the positive and negative helicity polarization tensors, and 
$k^ie^{(\pm)}_{ij}=0$. In a perfect fluid background, i.e. if there are no 
anisotropic stresses, both amplitudes satisfy the same wave equation,
\be\label{hwave}
\Box h= \ddot h + 2\frac{\dot a}{a}\dot h +k^2 h=0\ ,
\ee
where  $h  \equiv h_\pm$.
This equation of motion is obtained when expanding the gravitational action
in a Friedman universe to second order in $h$,
\be
S +\de S = -\frac{m_p^2}{2}\int d^4x\sqrt{-(g+\de g)}(R+\de R)~.
\ee
A brief calculation shows that the lowest (second order) contribution to 
$\de S$ can be written in Minkowski-space canonical form,
\be\label{canaction}
\de S^{(2)} = -\frac{1}{2}\int d^4x\left(\dd_\mu \phi\dd^\mu \phi-
    \frac{\ddot a}{a}\phi^2\right) = \int d^4x{\cal L}\,,
\ee
if we rescale $h$ as
 \be\label{rescaling}
h(\bx,t)={1\over \sqrt{2}\,m_p a(t)}\phi(\bx,t)\ .
\ee
Eq.~(\ref{canaction})  is the action of a canonical scalar field with 
time-dependent effective squared mass $m^2(t)= -\ddot a/a$. 
If the expansion of the Universe is slow enough (compared to the frequency 
of the mode under consideration), then the effective mass is 
negligible, and the theory describes a massless scalar field in Minkowski 
space,  which can be quantized according to the usual procedure: we first 
promote the field to an operator
\be\label{e:phihat}
 \hat\phi(\bx,t) =\int \frac{d^3k}{(2\pi)^3}\left[ e^{i\bk\cdot\bx}
 \chi_\bk(t)\hat b_\bk+ e^{-i\bk\cdot\bf{x}}
 \chi_\bk^*(t)\hat b_\bk^{\dagger}\right]\ ,
\ee
then we impose the commutation rules
\be\nonumber
[b_\bk,b^\dagger_{\bk'}]= (2\pi)^3\de^3(\bk-\bk')\ , \quad
 [b_\bk,b_{\bk'}]=[b^\dagger_\bk,b^\dagger_{\bk'}]=0 \,.
\ee
The field equations derived from the action (\ref{canaction}) 
lead to the mode equation
\be\label{modeq}
\ddot\chi_\bk +\left[k^2 -\frac{\ddot a}{a}\right]\chi_\bk = 0\,.
\ee

Within linearized gravity we can therefore quantize the metric fluctuations, 
provided the Universe expands adiabatically, by making use of the above 
rescaling of the amplitude $h$.

In particular, we now assume that the Universe is initially 
radiation-dominated, 
so that $a(t)=t$, and $\ddot a/a=0$, and $\phi$ represents exactly a 
massless scalar field in Minkowski space. We consider the vacuum initial conditions for
the modes $\chi_\bk(t)$ as given by
\be\label{chivac}
 \chi_\bk(t) =\frac{1}{\sqrt{2k}}e^{-ikt}\ .
\ee
More general initial conditions will be considered at the end of Sec.\ IV.
The field normalization is determined by the Klein-Gordon norm
\be\label{e:norm}
i\chi^*\stackrel{\leftrightarrow}{\dd_0}\chi
\equiv i(\chi^*\dd_0\chi -\chi\dd_0\chi^*) =1~.
\ee
Then, the field operator $\hat\phi$ and its canonically conjugate momentum, 
$\hat\Pi= \dd{\cal L}/\dd{(\dd_0\hat{\phi})}$ satisfy the canonical 
commutation relations.
The operators $\hat b_\bk$ define the $in$ vacuum by $\hat b_\bk|0_{\rm in}
\rangle =0 ~\forall \bk$.  In the following, this is the initial vacuum, 
void of particles by construction.

Suppose now that at a time $t=t_1$, the Universe changes abruptly from 
radiation-dominated to another expansion law. Then, the effective squared 
mass  no longer vanishes, and the initial modes, with $kt_1<1$ are amplified. 
Continuity 
requires that $\chi$ and $\dot \chi$ match at $t=t_1$, and these conditions 
determine the Bogoliubov coefficients, which relate the new modes 
$\chi_{\rm out}$ and operators $\hat b_{\bk \rm , out}$ to the old ones, 
$\chi_{\rm in}$ and $\hat b_{\bk \rm , in}$ by  \cite{BirDav}
\bea
\chi_{\rm out}(t)&=&\alpha \chi_{\rm in}(t)+\beta\chi^*_{\rm in}(t)\ ,
\\  \nonumber\\
\hat b_{\rm out}&=&\alpha^*\hat b_{\rm in}-\beta^*\hat 
    b^{\dagger}_{\rm in} \ .
\eea
With these relations, we can compute the number density of the
particles\footnote{We use the notion 'particle' is a somewhat 
sloppy way. These modes are particles in the standard sense of the term only 
once the mode has entered the horizon. Only then $N(k,t)$ is really a 
particle number, before it has rather to be related to the square
amplitude of field correlations. Of course there is no way of measuring
fluctuations with wavelengths larger than the horizon scale.}  created  
at the transition~\cite{BirDav}
\bea\label{e:Nkt1}
N(k,t_1)=\langle 0_{\rm in} |\hat b^{\dagger}_{\bk,\,\rm out} \hat b_{\bk,\,\rm out}
      | 0_{\rm in}\rangle=|\beta|^2\ .
\eea
Thus, the energy density $\rho\equiv \langle T^0{}_{0}\rangle$ can be written as
\bea\label{enden}
\rho_h={1\over a^4}\int {d^3k\over (2\pi)^3}kN(k,t_1)={1\over 2\pi^2a^4}
    \int dk k^3|\beta|^2\ ,
\eea
which implies the usual formula\footnote{Again, this is a physical graviton 
energy density only for scales well inside the horizon.}
\bea \label{e:rhoh}
{d\rho_h\over d\log k}\Big|_{\rm tot}={k^4|\beta|^2\over \pi^2 a^4}\ ,
\eea
where we have multiplied Eq.\ (\ref{enden}) by a factor 2 to take into 
account both 
polarizations. Note also that $k$ denotes comoving momenta/energy so that we 
had to divide by $a^4$ to arrive at the physical energy density. The second
quantity of interest is the power spectrum $P_h(k,t)$, defined by
\bea
4\pi \int {dk\over k}P_h(k,t)=\langle 0_{\rm in}| h(t,\bx)^2 |0_{\rm in}\rangle\ .
\eea
 Using Eqs.\ (\ref{rescaling}) and (\ref{e:phihat}), we obtain
\be\label{e:Ph1}
P_h(k,t)={k^2|\beta|^2\over (2\pi)^3 m_p^2 a^2}\ ,
\ee
where, again, we have multiplied by 2 to account for both polarizations.

Note that for all this it is not important that we consider a spin 2
graviton. The exactly same mode equation is obtained for a scalar field
and also for a fermion field. In the latter case, the commutation
relations have to be replaced by the corresponding anti-commutation relations.

\section{From radiation to matter era}\label{one}

Before discussing a transition from the radiation-dominated era to the
matter era, let us consider the transition from radiation to some 
generic power law expansion phase, $a\propto t^q$ with $q\neq 1$ at some
time $t_1$. In the new era $\ddot a/a = q(q-1)/t^2 \neq 0$. Note that only 
if $q>1$ (or if $q<0$ which corresponds to inflation or contraction), so 
that $m^2(t) =-q(q-1)/t^2$ is negative, we will have significant 
particle production. Since the expansion law is related to the equation of 
state parameter $w=P/\rho$ via~\cite{myBook}
\be\label{e:qw}
q =\frac{2}{1+3w}\,,
\ee
this requires $w<1/3$.

Let us start in the vacuum during the radiation era, then $\chi$ is given 
entirely by the negative frequency modes, Eq.~(\ref{chivac}). This means that
we consider the situation where there are no significant gravity waves 
present from an earlier inflationary epoch. Here, we really want to study 
the production due solely to the radiation/matter transition. 
The general solution of the mode equation~(\ref{modeq}) in the new era 
are the spherical Hankel functions~\cite{AS} of order $q-1$,
\bea\label{e:solq}
\chi_\bk(t)&=&\frac{\al_1}{\sqrt{2k}}zh_{q-1}^{(2)}(z) +
\frac{\beta_1}{\sqrt{2k}}zh_{q-1}^{(1)}(z)  
\eea
where $z=kt$. Note that inside the horizon, i.e. for $z\gg 1$, 
$zh^{(2)}_{q-1}(z) \propto \exp(-iz)$ corresponds to the negative frequency 
modes while $zh^{(1)}_{q-1}(z) \propto \exp(iz)$ corresponds to positive 
frequency modes. We match $\chi$ and $\dot\chi$ at $t=t_1$ to the 
radiation-dominated vacuum solution (\ref{chivac}). A brief calculation 
yields the coefficients ($z_1\equiv kt_1$)
\bea\label{e:aq1}
\hspace*{-5mm} \al_1 &=& -\frac{i}{2}e^{-iz_1}\left[(iz_1+q)h^{(1)}_{q-1}(z_1) -
              z_1h_q^{(1)}(z_1)\right]\,, \\   \label{e:bq1}
\hspace*{-5mm} \beta_1 &=&-\frac{i}{2}e^{-iz_1}\left[(iz_1+q)h^{(2)}_{q-1}(z_1) -
              z_1h_q^{(2)}(z_1)\right]\, .
\eea
This instantaneous matching condition is good enough for frequencies for 
which the transition is rapid, i.e. $z_1\ll 1$. In fact, for frequencies with 
$z_1>1$, the transition is adiabatic and no particle creation will take place. 
This can also been seen when considering the limits of the above result 
for large $z_1$. Then $\al_1\ra 1$ and $\beta_1\ra 0$, but strictly speaking 
the above approximations are not valid in this regime where no particle 
creation takes place.
We therefore concentrate on $z_1\ll 1$.

Let us now study the specific case of the radiation--matter transition, 
i.e. $q=2$. Then we have to consider spherical Hankel functions of 
order 1 and the solution is given by
\be
\label{solmat}
\chi_\bk(t) = \frac{\al_1}{\sqrt{2k}}\frac{z-i}{z}e^{-iz} +
\frac{\beta_1}{\sqrt{2k}}\frac{z+i}{z}e^{iz} \,.
\ee
The matching at $t_1$ now yields
\be\label{e:ab1}
\al_1 =1 +\frac{i}{z_1} -\frac{1}{2z_1^2} \, , \qquad
\beta_1 =-\frac{1}{2z_1^2}\,e^{-2iz_1}~.
\ee

We want to evaluate the quantum field $\hat\phi$ at late time,
when $z\gg1$ and the mode $k$ under consideration is sub-horizon. Then,
the solution (\ref{solmat}) is again the Minkowski solution,
\be\label{solmatin}
\chi_\bk(t)\simeq \frac{\al_1}{\sqrt{2k}}e^{-iz} +
\frac{\beta_1}{\sqrt{2k}}e^{iz}~.
\ee

The number of gravitons generated during the matter era
(before $z\gg 1$) is, see Eq.~(\ref{e:Nkt1})
\bea
N(k,t) &=&  |\beta_1|^2\,.
\eea

The graviton power spectrum is given by Eq.~(\ref{e:Ph1}), and
the energy density by Eq.~(\ref{enden}). 

Using that $\rho_\mr{rad}a^4 \equiv \rho_\mr{rad}(t_1)a_1^4 =
\frac{3}{2}m_p^2\HH_1^2a_1^2$ and $|\beta_1|^2 =z_1^{-4}/4$, we obtain
\bea\nonumber
\frac{d\Om_h(k)}{d\log k} &=& \frac{2\Om_{\mr{rad}}}{3\pi^2}
\frac{k^4}{m_p^2\HH_1^2a_1^2}|\beta_1|^2 =
 \frac{ \Om_{\mr{rad}}}{6\pi^2}\left(\frac{H_1}{m_p}\right)^2 \\
 &=& \Om_{\mr{rad}}\frac{g_\mr{eff}}{36\times 30}
  \left(\frac{T_1}{m_p}\right)^4 ~.    \label{e:res1}
\eea
For the second equal sign we have used that $\HH_1=1/t_1 =a_1H_1$, which
is strictly true only in the radiation era, in the matter era we have
 $\HH=2/t$ and at the transition a value between $1$ and $2$ would probably 
be more accurate. But within our approximation of an instant transition,
we do not bother about such factors. For the last equal sign we used
$H_1^2 =\rho_\mr{rad}/(6m_p^2)$ with
$\rho_\mr{rad} =g_\mr{eff}\frac{\pi^2}{30}T^4$ where $g_\mr{eff} =
N_B +\frac{7}{8}N_F$ is the effective number of degrees of freedom.

For a generic transition we obtain $|\beta_1|^2 =z_1^{-2q}/4$ so that
\be
\frac{d\Om_h(k)}{d\log k} \simeq \Om_{\mr{rad}} 
  \left(\frac{T_1}{m_p}\right)^4z_1^{4-2q} ~.    \label{e:res1gen}
\ee
This spectrum is blue (i.e. growing with $k$) if $q<2$ and red otherwise.

As $z_1<1$ in the regime of validity of our formula, we need a red spectrum
i.e. $q>2$ to enhance the gravitational wave energy density with respect 
to the result from the radiation $\ra$ matter transition. According to 
Eq.~(\ref{e:qw}), this requires $-1/3<w<0$, a slightly negative pressure, 
but still non-inflationary expansion.

In the standard radiation to matter transition when three species of
left handed neutrinos and the photon are the only relativistic degrees
of freedom, we have $g_\mr{eff}=29/4$. For this transition
$$T_1=T_\mr{eq}\simeq 0.85\mr{eV} \simeq 0.35\times 10^{-27}m_p\,,$$ 
hence the result (\ref{e:res1}) is completely negligible.

\section{More than one radiation-matter transition}\label{more}

We now consider an early matter dominated era. At some high temperature 
$T_1\gg T_\mr{eq}$, corresponding to a comoving time $t_1$, a massive 
particle may start to dominate the Universe and render it matter-dominated. 
At some later time $t_2$, corresponding to temperature $T_2$, this
massive particle decays and the Universe becomes radiation-dominated again,
until the usual radiation--matter transition, which takes place at
$ T_\mr{eq}\equiv T_3$. We want to determine the gravitational wave spectrum
and the spectral density parameter $d\Om_h/d\log(k)$ as functions of $T_1$ 
and $T_2$.

Let us first again start with the vacuum state in the radiation eta before 
$t_1$. When,  we just obtain the results (\ref{e:ab1}) for the Bogoliubov 
coefficients $\al_1$ and $\beta_1$ after the first transition.
To evaluate the matching conditions at the second transition, matter to
radiation, we set
$$\chi =\frac{\al_2}{\sqrt{2k}}e^{-ikt} +\frac{\beta_2}{\sqrt{2k}}e^{ikt}
 \,, \quad t\ge t_2\,.$$
Matching $\chi$ and $\dot\chi$ at $t_2$ we can relate the new 
coefficients $\al_2$ and $\beta_2$ to $\al_1$ and $\beta_1$. A brief 
calculation gives
\bea
\al_2 &=&\al_1\left(1-\frac{i}{z_2} -\frac{1}{2z_2^2}\right) +
   \frac{\beta_1}{2z_2^2}e^{2iz_2}  \nonumber \\ \label{e:ab2}
 &=& \al_1f(z_2) +\beta_1 g(z_2) \, , \\
\beta_2 &=& \beta_1\left(1+\frac{i}{z_2} -\frac{1}{2z_2^2}\right)  +
     \frac{\al_1}{2z_2^2}e^{-2iz_2} \\
 &=& \beta_1\bar{f}(z_2) +\al_1 \bar{g}(z_2) ~,
\eea
or in matrix notation
\bea
 \left(\begin{array}{c} \al_2 \\ \beta_2\end{array}\right) &=& 
 {M}(z_2)  \left(\begin{array}{c} \al_1 \\ \beta_1\end{array}\right)\,,
 \qquad \mbox{with} \\   \label{e:M}
   {M}(z) &=& \left(\begin{array}{cc} f(z) & g(z) \\ \bar{g}(z) & 
\bar{f}(z) \end{array}\right) \\
 {M}^{-1}(z) &=& \left(\begin{array}{cc} \bar{f}(z) & -g(z) \\ 
 -\bar{g}(z) & f(z) \end{array}\right)\ .
\eea
The fact that $M\in Sl(2,\cx)$, i.e., $|f(z)|^2 -|g(z)|^2=1$ ensures 
that the normalization condition (\ref{e:norm})
which translates to the condition $|\al|^2 -|\beta|^2=1$ for the Bogolioubov
coefficients of a free field, is maintained at the transition.
Finally, the matching at the usual radiation--matter transition yields
\bea
\al_3 &=&\al_2\left(1+\frac{i}{z_3} -\frac{1}{2z_3^2}\right)
     - \frac{\beta_2}{2z_3^2}\,e^{2iz_3}~, \nonumber \\ \label{e:ab3}
\beta_3 &=&\beta_2\left(1-\frac{i}{z_3} -\frac{1}{2z_3^2}\right)
   -\frac{\al_2}{2z_3^2}\,e^{-2iz_3}~,\\
\left(\begin{array}{c} \al_3 \\ \beta_3\end{array}\right) &=& 
 M^{-1}(z_3)  \left(\begin{array}{c} \al_2 \\ \beta_2\end{array}\right)\\
\left(\begin{array}{c} \al_3 \\ \beta_3\end{array}\right) &=& 
 M^{-1}(z_3)M(z_2) M^{-1}(z_1)\left(\begin{array}{c}1 \\ 0\end{array}\right)  \,.
\eea

\begin{figure}[h]
\includegraphics[width=70mm]{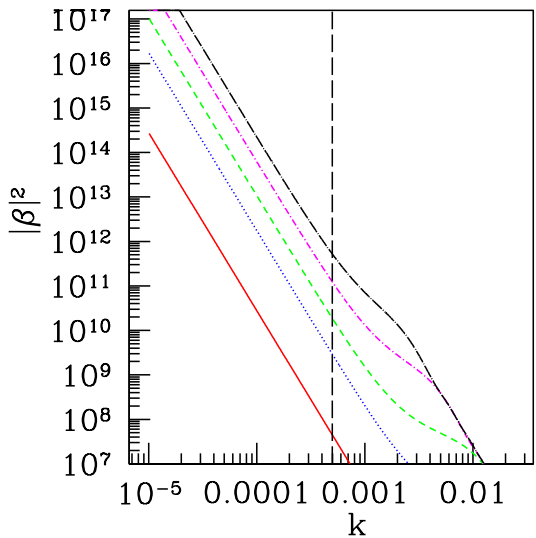}
\caption{\label{f:beta} The Bogoliubov coefficient $|\beta_3(k)|^2$
for $t_1=1$, and $t_3=2000$, with various values of $t_2$. Only the k-values 
left of the vertical dashed line satisfy $kt_3<1$. They show clearly a 
$k^{-4}$ slope and the amplitude is well approximated by (\ref{e:ab3<1}).}
\end{figure}

To obtain the power spectrum and energy density in this case, we simply
have to replace $|\beta|^2$ in Eqs.~(\ref{e:Ph1}) and (\ref{e:rhoh}) by
 $|\beta_3|^2$. In Fig.~\ref{f:beta} we plot $|\beta_3|^2$ as a function
 of $z_3$ for different choices of $t_2$. The instantaneous transition 
approximation breaks down for $z_3>1$, hence only the left side of the 
vertical line is physical. For the right side one would have to solve the 
mode equation numerically, but since we know that particle production is 
suppressed for these frequencies, we do not consider them. We concentrate 
on $z_3\le 1$. For these wave numbers, also $z_1<z_2<z_3<1$.

 This allows the following approximations,
 \bea\label{lowordercorr}
&& \al_1 \simeq  \beta_1 \simeq -\frac{1}{2z_1^2} \,, \quad z_1\ll 1 \\
\label{e:ab2<1}
&& \al_2 \simeq -\beta_2 \simeq \frac{iz_2}{3z_1^2}\,, \quad z_2\ll 1 \\
\label{e:ab3<1}
&& \al_3 \simeq \beta_3 \simeq  -\frac{2z_2}{3z_1^2z_3}\,,
\quad z_3\ll 1\,. 
\eea
To obtain the results (\ref{e:ab2<1}) and (\ref{e:ab3<1}) we have to expand 
the exact expression (\ref{e:ab2}) to fourth order and (\ref{e:ab3}) to 
second order, but we consider only the largest term in the 
result given above, using also $z_1 < z_2<z_3$. Therefore,
in the approximate expression (\ref{e:ab2<1}), where we have neglected a term 
proportional to $z_1/z_2^2$, one no longer sees that
$\al_2\ra 1$ and $\beta_2\ra 0$ when $t_2\ra t_1$ and hence $z_2\ra z_1$.
In this case there is no intermediate matter-dominated era and therefore
no particle creation, hence $\beta_2=0$. This can be seen from the exact 
expression given in Eq.~(\ref{e:ab2}).

Within these approximations, Eqs.~(\ref{e:Ph1}) and (\ref{e:rhoh}) lead to
\bea
P_h(k) &=& \frac{1}{(2\pi)^3}\left(\frac{t_2}{m_pat_1^2t_3k}\right)^{2} \,,
\quad kt_3<1  \\
\frac{d\Om_h}{d\log k} &=& \frac{\Om_\mr{rad}}{\pi^2}
   \frac{k^4|\beta_3|^2}{a_1^4\rho_\mr{rad}(t_1)}  \nonumber \\  &=&
   \Om_\mr{rad}\frac{g_\mr{eff}(T_1)}{18\times 45}\left(\frac{T_1}{m_p}\right)^4
   \left(\frac{T_\mr{eq}}{T_2}\right)^{2}  \,,~ kt_\mr{eq}<1\,.  \nonumber \\  
   && \label{e:res2}
\eea
This result can be generalized to several, say $N$, intermediate radiation 
$\ra$ matter transitions at times $t_{2n-1}$ and back to radiation at time 
$t_{2n}$, $1\le n\le N$, with the result
\be
|\beta_{2N+1}|^2 \simeq \frac{1}{(kt_1)^4}\left(
   \frac{T_3 \cdots T_{2N-1}T_\mr{eq}}{T_2\cdots T_{2N}}\right)^2 \,.
\ee

Hence, each return to the radiation-dominated era at some intermediate 
temperature $T_{2n}$ leads to a suppression factor $(T_{2n+1}/T_{2n})^2$, 
where $T_{2n+1}$ denotes the temperature at the start of the next matter era.

On large scales, $kt_\mr{eq}<1$, the energy density spectrum is flat. The 
best constraints on an intermediate radiation-dominated era therefore come
from the largest scales, i.e. from observations of the cosmic microwave 
background (CMB) as we shall discuss in the next section.

We now briefly consider the case when the initial conditions differ 
from the vacuum case, Eq.\ (\ref{chivac}). We assume an arbitrary initial 
state of the field given by
\be\label{chivac2}
 \chi_\bk(t) =\frac{\alpha_0}{\sqrt{2k}}e^{-ikt}+
\frac{\beta_0}{\sqrt{2k}}e^{ikt}\ ,
\ee
together with the normalization condition which ensures that the field is 
canonically normalized, $|\alpha_0|^2-|\beta_0|^2=1$. The same calculations 
as above now yield
\bea
\left(\begin{array}{c} \al_3 \\ \beta_3\end{array}\right) &=& 
 M^{-1}(z_3){M}(z_2) M^{-1}(z_1)  \left(\begin{array}{c} \al_0 \\
 \beta_0\end{array}\right)\,,
\eea
where $M(z)$ is the matrix giving the transition from matter to radiation 
defined in Eq.~(\ref{e:M}).

Expanding this in  $z_1$, $z_2$ and $z_3$, using $z_1<z_2<z_3<1$ one finds 
that to lowest non-vanishing order, the final result for $\beta_3$ depends 
only $|\alpha_0+\beta_0|$. However, if the phase of 
$\alpha_0$ and $\beta_0$ are nearly opposite, i.e., $\alpha_0 \simeq -\beta_0$ ,
and if $|\alpha_0|$ and therefore also 
$|\beta_0|$ are much larger than $1$, a correction proportional 
to $|\alpha_0-\beta_0|$ becomes important.
More precisely, the last of Eqs.\ (\ref{lowordercorr}) now is replaced by
\be\label{e:novac}
\beta_3\simeq {2z_2\over 3 z_1z_3}\left[-\frac{1}{z_1}(\alpha_0+\beta_0)+
    2i(\alpha_0-\beta_0)\right]\ .
\ee
If $\al_0=1$ and $\beta_0=0$, the second term can be neglected with respect 
to the first one and we reproduce the previous result (\ref{e:ab3<1}).
As we see from this equation, a large phase difference 
between $\alpha_0$ and $\beta_0$ changes not only the amplitude but also 
the slope  of the spectrum. Of course in concrete examples, like for
a previous inflationary period, see Ref.~\cite{VS2}, the coefficients
$\al_0$ and $\beta_0$ also depend on the wavenumber.

\begin{figure}[ht]
\includegraphics[width=70mm]{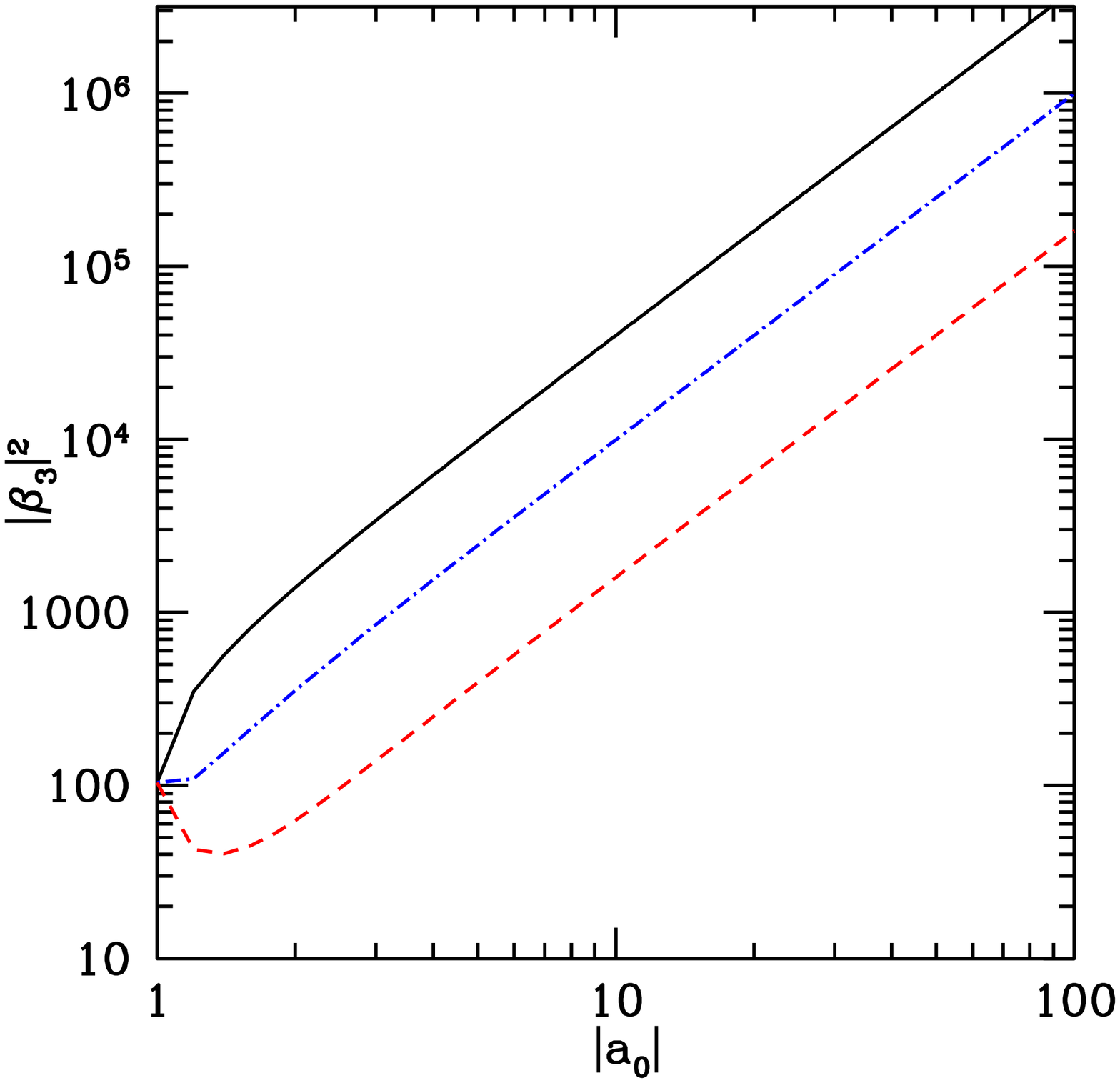}\\
\includegraphics[width=70mm]{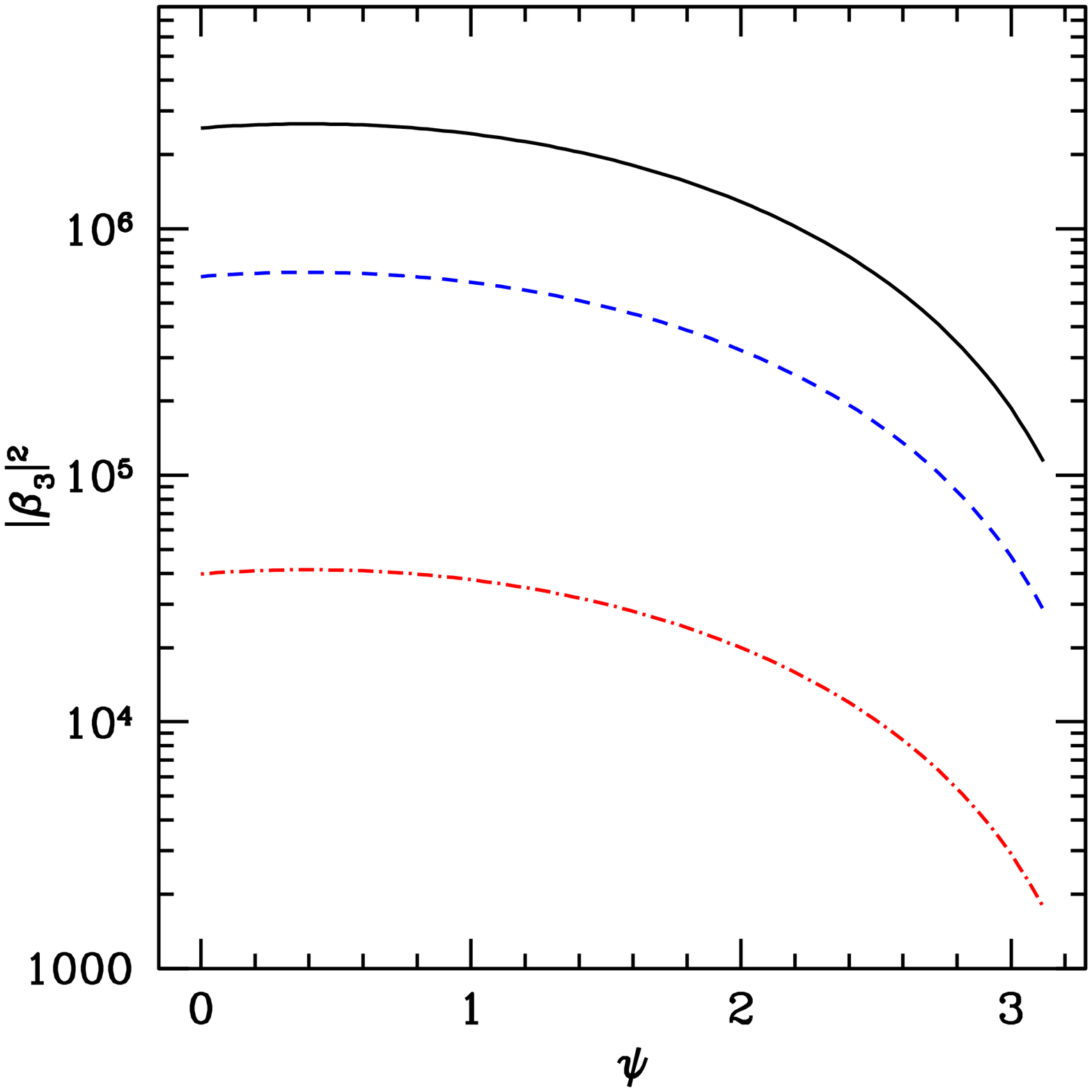}
\caption{\label{f:novac}In the top panel we show $|\beta_3|^2$ as a function
of $\al_0$ for $\al_0$ and $\beta_0$ in phase (top, solid, black line), of
opposite phase (lowest, dashed, red line) and with a phase difference of 
$0.8\pi$ (middle, dot-dashed, blue line).\\
In the bottom panel we show   $|\beta_3|^2$ as a function of the phase 
difference $\psi$ for $|\al_0|=80$ (top line)  $|\al_0|=40$ (middle line) and 
$|\al_0|=10$ (lowest line). We have chosen $z_1 =0.1$ in this plot and 
the overall vertical normalization is in units of 
$\left(2z_2/(z_1z_3)\right)^2$.
}
\end{figure}

In Fig.~\ref{f:novac} we show the dependence of $|\beta_3|^2$  on $|\al_0|$
for different values of the relative phase between $\al_0$ and $\beta_0$ 
(top panel) and as a function of the relative phase for different values of
$|\al_0|$. The difference of  $|\beta_3|^2$ between the case where  
$\al_0$ and $\beta_0$ are perfectly in phase and of opposite phase is of 
the order of $1/z_1$, if $|a_0|$ is significantly larger than $1$. This 
is already evident from Eq.~(\ref{e:novac}). In the Fig.~\ref{f:novac} we
have chosen $z_1=0.1$, a unrealistically high value, in order to have a 
better visibility of the phase dependence which then changes 
$|\beta_3|^2$ only by one order of magnitude.

In conclusion, in the case of a non-vacuum initial state, $|\al_0|$ 
significantly larger than $1$, graviton 
production is enhanced typically by a factor of order $|\al_0|^2 \sim 
|\beta_0|^2$ which is the number of initial particles. Hence in addition to 
the spontaneous creation we now also have induced particle creation which 
is proportional to the initial particle number and much larger than 
the spontaneous creation if the particle number is large.
An interesting point is that the phase shift between $\al_0$ and $\beta_0$ 
can significantly affect the final spectrum.

\section{Discussion and Conclusions}\label{conc}
The fact that the observed CMB anisotropies are of the order of $10^{-5}$
yields  a strong limit on gravitational waves with wave numbers of the order
of the present Hubble scale, see e.g.~\cite{bb}.
\be\label{e:lim}
\left.\frac{d\Om_h}{d\log k}\right|_{k=H_0} < 10^{-15} \,.
\ee
With $\Om_\mr{rad} \simeq 10^{-5}$, and Eq.\ (\ref{e:res2}), this implies
the limit
\be\label{e:cond}
\left(\frac{T_1}{m_p}\right)^2\frac{T_\mr{eq}}{T_2} <   10^{-5}\,.
\ee
 We know that during nucleosynthesis the Universe was radiation-dominated, 
hence $T_2 \ge 0.1$MeV. With $T_\mr{eq}\sim 1$eV, the above 
inequality reduces to $T_1<m_p$ for the value $T_2\simeq 0.1$MeV, and it is
even less stringent for higher values of $T_2$. Hence even though the 
production of gravitons during an intermediate matter era is of 
principal interest, we cannot derive stringent limits on $T_1$ and $T_2$. 
On the other hand, for values of $T_1$ and $T_2$ 
close to the maximal respectively minimal value, $T_1\sim m_p$ and 
$T_2\sim 0.1$MeV, these gravitons would leave a detectable signature in 
the cosmic microwave background.

We can, however, use this effect to limit any intermediate era with $q>2$,
i.e. $-1/3<w=P/\rho<0$. According to
Eq.~(\ref{e:res1gen}), in the general case, the particle number is of the 
order of $|\beta_1|^2 \simeq z_1^{-2q}$, so that
\be\label{e:res2gen}
\frac{d\Om_h}{d\log k}  \simeq \Om_\mr{rad}(kt_1)^{4-2q}
\left(\frac{T_1}{m_p}\right)^4  \left(\frac{T_\mr{eq}}{T_2}\right)^2\ ,\quad 
  z_1\ll 1\,.
\ee
As above, $T_2$  denotes the temperature at which the Universe returns 
to the radiation-dominated state, hence nucleosynthesis requires $T_2>0.1$MeV. 
In this case, if $q>2$, the spectrum becomes red and, at $k\sim H_0$, the 
limit can become quite interesting. Hence, graviton production can 
significantly limit a (non-inflationary) phase with negative pressure in 
the early universe. For $k=H_0$, using $t_1=\HH_1^{-1} =1/(a_1H_1)$, we obtain
$$ (H_0t_1)^{-1} \simeq \left(\frac{g_\mr{eff}(T_1)}{10^5}\right)^{1/2}
   \left(\frac{T_1}{T_0}\right)^{3}\,,$$
where $T_0\simeq 0.2\times 10^{-4}$eV is the present temperature of the 
Universe.
This can be a significant factor for large values of $T_1$.
Inserting this expression in Eq.~(\ref{e:res2gen}), the limit~(\ref{e:lim}) 
yields
\be
\left(\frac{g_\mr{eff}(T_1)}{10^5}\right)^{q-2}
  \left(\frac{T_1}{T_0}\right)^{6(q-2)} \left(\frac{T_1}{m_p}\right)^4
  \left(\frac{T_\mr{eq}}{T_2}\right)^2 \lsim 10^{-10} ~.
\ee
For example, for $w=-1/21$, ($q=7/3$), we already obtain
\be
 \frac{T_1}{T_2}\left(\frac{T_1}{m_p}\right)^2 \lsim 10^{-9}\,.
 \ee
E.g. for $T_2=1$MeV this implies $T_1<10^8$GeV.
 For smaller values of $w$ the limit becomes more stringent.

In this paper we have shown that there is cosmological particle production
of massless modes whenever the expansion law is not radiation-dominated
so that $\ddot a/a \neq 0$. This term acts like a time-dependent mass
and leads to the production of modes with comoving energy
$k^2<|\ddot a/a | \simeq \HH^2$, hence with physical energy $\om =k/a\lsim H$.
One readily sees that particle production is significant only if $q<1$, i.e. 
the squared mass $ -\ddot a/a <0$.
 When starting from a radiation-dominated Universe, we do have
a well defined initial vacuum state also for the super-horizon modes which 
can be amplified by a transition to another expansion law.
We have also given the expressions for the produced particle number in the 
case of a generic, non-vacuum initial state, Eq.~(\ref{e:novac}).
This can be applied to an arbitrary inflationary, pre-big bang or bouncing 
model, which may already contain gravitons before the first radiation era.

We have explicitly calculated the production of gravitons for a vacuum 
initial state and have arrived at the following main conclusions:
\begin{itemize}
\item[i)~] The gravitons produced after the standard
 radiation--matter transition are completely negligible.
\item[ii)] If we introduce a matter-dominated era in the early universe,
this  leads to a flat energy spectrum of gravity waves which, in the most
optimistic case, can be sufficient to contribute to the CMB tensor 
anisotropies in an observable way.
\item[iii)] A phase of expansion with $-1/3<w<0$ leads to a red spectrum 
of gravitons. Such a phase in the early universe is severely constrained
mainly by the amplitude of CMB anisotropies.
\item[iv)] A graviton spectrum present at the beginning of the radiation 
era can become significantly amplified and modified by intermediate 
not-standard evolution of the universe. 
\end{itemize}

\noindent
{\bf Acknowledgment:} We thank John Barrow for valuable comments.
This work is supported by the Swiss National Science Foundation.

\end{document}